\documentclass[a4paper]{article}

\usepackage[top=2.8cm,bottom=3cm,left=3cm,right=3cm]{geometry}
\usepackage{graphicx}
\usepackage{txfonts}
\usepackage{amsfonts}
\usepackage{amssymb}
\usepackage{booktabs}
\usepackage{xcolor}
\usepackage{lineno}
\usepackage{natbib}

\begin{document}


\Large
\textbf{\begin{center}{Artificial Intelligence Could Have Predicted All Space Weather Events Associated with the May 2024 Superstorm} \end{center}}

\normalsize

Sabrina Guastavino$^{1,2}$, Edoardo Legnaro$^{1}$, Anna Maria Massone$^{1}$, and Michele Piana$^{1,2}$ \\

\hspace{-0.5cm}$^1$ MIDA, Dipartimento di Matematica, Universit\`a di Genova, via Dodecaneso 35 16146 Genova, Italy  \\
$^2$ Istituto Nazionale di Astrofisica, Osservatorio Astrofisico di Torino, via Osservatorio 20 10025 Pino Torinese Italy \\


\begin{center}
\textbf{Abstract}
\end{center}
Space weather, driven by solar flares and Coronal Mass Ejections (CMEs), poses significant risks to technological systems. Accurately forecasting these events and their impact on Earth's magnetosphere remains a challenge because of the complexity of solar-terrestrial interactions. This study applied artificial intelligence (AI) to predict the chain of events associated with the May $2024$ superstorm, including solar flares from NOAA active region 13644, Earth-directed CMEs, and a violent geomagnetic storm. Using magnetogram cut-outs, a Vision Transformer was able to classify the evolution of the active region morphologies, and a video-based deep learning method predicted the occurrence of solar flares; a physics-driven model improved the precision of CME travel-time prediction using coronal observations and solar wind measurements; and a data-driven method exploited these in situ measurements to sound alerts of the geomagnetic storm unrolled over time. The results showed unprecedented accuracy in predicting CME arrival with uncertainty as small as one minute. Moreover, these AI models outperformed traditional methods in predicting solar flares occurrences, onset, and recovery phases of the geomagnetic storm. These findings highlight the impressive potential of AI for space weather forecasting and as a tool to mitigate the impact of extreme solar events on critical infrastructure.
 \\

\textbf{keywords:} Physical Sciences; Astronomy and Planetary Science (or Physics); Space Physics; Solar Physics

\section{Introduction}
\label{sec:introduction}
In May $2024$, a series of extreme solar events, including a class X$8.7$ solar flare and multiple interacting CMEs, resulted in the second most severe geomagnetic storm of the space exploration era \citep{diaz2024monitoring,kwak2024observational}. This impressive storm has once again brought to the attention of scientists, technologists, space stakeholders, and policy makers the need for computational tool able to predict and characterize events associated with space weather. Indeed, this study of solar activity and its interaction with the Earth’s magnetosphere is a critical research field due to its far-reaching impact on technological systems and human infrastructure \citep{2007LRSP....4....1P,2010hssr.book.....S}. Solar flares and Coronal Mass Ejections (CMEs) are the most significant drivers of space weather. Specifically, solar flares are intense bursts of electromagnetic radiation emitted by the Sun \citep{tandberg1988physics,piana2022hard}, while CMEs involve the ejection of large volumes of plasma and magnetic fields from the solar corona into space \citep{2012LRSP....9....3W,georgoulis2019source}. 
When directed toward Earth, CMEs can cause significant geomagnetic disturbances, making accurate forecasting of the solar flare onset and CME travel times crucial for preparedness and mitigation efforts \citep{2001JGR...10629207G,morosan2019multiple,lara2020space,shen2022propagation}. Additionally, forecasting the geomagnetic storm intensity caused by CMEs upon their arrival is equally important, as the severity of these storms can vary based on the properties of the CME and the interaction with the Earth’s magnetic field \citep{gosling1990coronal,2007JGRA..11210102Z,gopalswamy2016history,lugaz2017interaction,nitta2021understanding}.

Despite advances in observational capabilities and model development, so far there remains a substantial uncertainty in both flare forecasting and CME travel time predictions, with the latter averagely amounting to approximately $12$ hours. This uncertainty poses challenges for operational space weather forecasting \citep{2018SSRv..214...21R}, as it reduces the window for preventive measures to be taken, particularly when mitigating the potential impacts on critical infrastructures and space assets \citep{2017RiskA..37..206E}. Additionally, the current empirical and deterministic models developed to address the prediction of space weather phenomena often fail to account for the full complexity of solar-terrestrial interactions, which can lead to inaccuracies in the prediction of the severity of the resulting geomagnetic storms \citep{2015SoPh..290.1775M}.

In recent years, artificial intelligence (AI) has emerged as a promising approach to enhance the predictive capabilities of space weather models \citep{camporeale2018machine,2019SpWea..17.1166C,gombosi2021sustained,goel2023role,2023SpWea..2103514N}. Indeed, AI-based methods have the potential to overcome some of the limitations of traditional models by leveraging large datasets and learning complex patterns that may not be easily captured by empirical or deterministic approaches \citep{2018SoPh..293...28F,murray2018importance,zhelavskaya2019systematic,campi2019feature,cicogna2021flare,guastavino2023operational,2024ApJ...971...94G}. In particular, in the last two years, a battery of very sophisticated, new generation AI tools have been developed to exploit both remote sensing and in-situ measurements, with the aim of characterizing and predicting space weather events with unprecedented reliability and accuracy. For example, as far as active region (AR) classification is concerned, \citet{legnaro2024deep} explored the most recent advancements in deep learning architectures for image processing when applied to the classification of both continuous images and magnetograms recorded by the Helioseismic and Magnetic Imager \citep[HMI;][]{2012SoPh..275..207S} onboard the Solar Dynamics Observatory \citep[SDO;][]{2012SoPh..275....3P}. 
This study showed that with a robust training process, including data augmentation and transfer learning, Vision Transformers (ViTs) and Convolutional Neural Networks (CNNs) systematically reach robust performances with impressively high stratification scores. 

As far as flare forecasting is concerned, based on videos of HMI magnetograms, \citet{guastavino2022implementation} developed a Long-term Recurrent Convolutional Network (LRCN) made of a parallel architecture of CNNs and of a Long Short Term Memory (LSTM) algorithm for predicting flares above GOES class C or M. This approach proved the possibility to bypass the need for a priori feature extraction from magnetograms, which is typical of standard machine learning approaches \citep{georgoulis2021flare}.

Regarding CME characterization, \citet{2023ApJ...954..151G} realized the pioneering attempt to incorporate physics-based constraints into neural networks for predicting CME travel times. By combining the predictive power of machine learning with a solid foundation in physics represented by the aerodynamic drag \citep{2013SoPh..285..295V,napoletano2018probabilistic}, the model was able to reduce the uncertainties associated with CME travel time predictions by approximately $25\%$. 

When it comes to predicting the severity of geomagnetic storms, particularly in the context of CME-driven storms, \citet{telloni2023prediction} showed how AI models can analyze solar wind data to predict the occurrence of such storms. Further, by evaluating key predictive features and ranking their importance \citep{camattari2024classifier}, the AI approach employed in \cite{2024ApJ...971...94G} led to more accurate predictions of the severe Earth's magnetospheric disturbances.

Building on these successes, this paper shows that AI-driven methodologies could have predicted the entire chain of space weather events that characterized the May 2024 solar storm, which was so far the most impressive and violent solar storm of the present solar cycle. These events provided a unique case study for testing AI-driven predictive models. The primary aim of our work was to utilize AI tools to predict the onset and evolution of the solar flares triggering the storm, progressing through the CME transit times, and culminating in the prediction of the geomagnetic storm. The overall results of this analysis are unprecedentedly accurate forecasts with significantly reduced uncertainties with respect to traditional methods. In particular, it has been evaluated the ability of physics- and data-driven AI models to predict CME travel times and assess the intensity of the induced geomagnetic storms, thereby offering a comprehensive tool for operational space weather forecasting.


\section{Results}

\subsection{AR13664 and its consequences}

The May $2024$ superstorm \citep{kwak2024observational,hayakawa2024solar} began with the emergence of the unusually active region AR13664 in early May $2024$, which started releasing a sequence of increasingly strong solar flares, the most powerful of which was a massive X$8.7$-class event on May $14$, one of the most intense flares recorded in recent history. Between May 7 – 11, multiple strong solar flares, with the strongest peaking with a rating of X5.8, initiated a series of CMEs that were directed toward Earth, interacting with each other in the interplanetary medium and merging into a single, highly energetic, and fast-moving plasma cloud. This interaction resulted in an exceptionally fast CME with a velocity exceeding $1000$ km$/$s, reaching Earth in less than $2$ days, and generating a geomagnetic storm classified as G$5$ on the National Oceanic and Atmospheric Administration (NOAA) scale. Such a short travel time significantly reduced the window of opportunity for space weather forecasters and preparedness measures on Earth.

\begin{figure}
    \centering
    \includegraphics[width=0.9\linewidth]{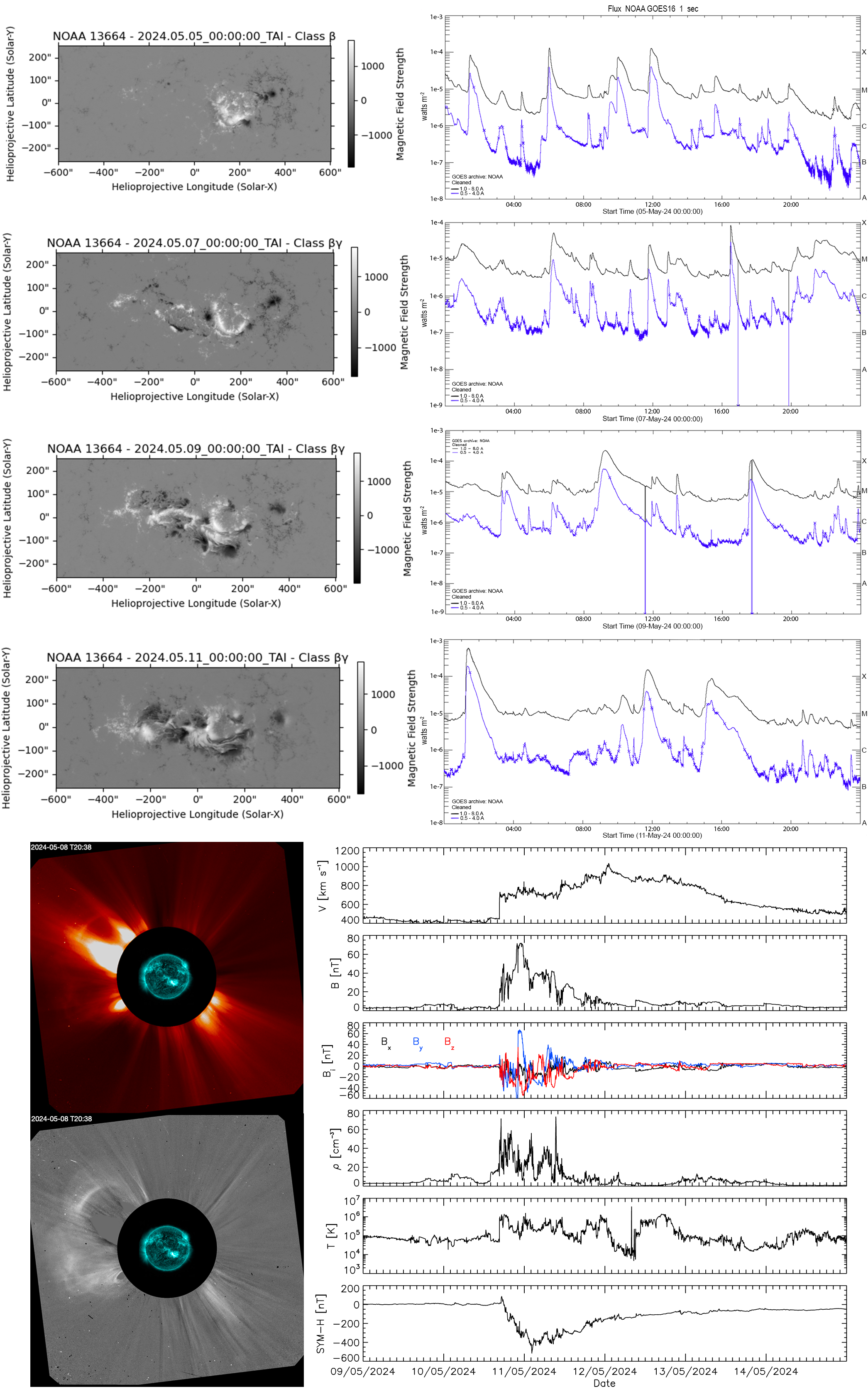}
    \caption{Remote sensing observations and in-situ data utilized for prediction. This figure shows magnetogram cutouts of NOAA Active Region 13664, accompanied by plots illustrating variations in magnetic field strength over time for each of the four dates. At the bottom, observations of CMEs in different wavelengths are displayed, along with plots of various solar wind and space weather parameters, including speed, density, temperature, and magnetic field components.}
    \label{fig:data}
\end{figure}

AR13664 entered the HMI's field of view around 2024-05-02 and exited around 2024-05-13 (see Figure \ref{fig:data}, the four top left panels). GOES began sounding alarms associated to two M-class flares on May 5 and kept on revealing more and more events even after the active region exited the HMI field of view, with just a one-day duration pause on May 5 (see Figure \ref{fig:data}, the four top right panels). The shock observed locally at L$1$ at $16$$:$$36$ on May $10$, $2024$ was potentially associated with $5$ CMEs emitted in rapid succession, as also reported on the Community Coordinated Modeling Center (CCMC) CME Scoreboard (https://kauai.ccmc.gsfc.nasa.gov/CMEscoreboard/). 
According to magnetohydrodynamic (MHD) Enlil simulations \citep{2003AdSpR..32..497O}, the last of these CMEs, shown in the last two bottom left panels of Figure \ref{fig:data} and which appeared in the field of view of the Large Angle and Spectrometric COronagraph (LASCO) \citep{brueckner1995large} at $19$$:$$12$ on May $8$, $2024$, would have cannibalized the previous ones during its journey to Earth, merging into a single highly energetic, fast-moving magnetic cloud. The arrival at L$1$ of the magnetic cloud was readily identified from the rotation of the interplanetary magnetic field associated with the onset of the geomagnetic storm. From the in-situ data (Figure \ref{fig:data}, the last six bottom right panels) it is apparent that the magnetic cloud arrived at Earth at $18$$:$$15$ on May $10$, $2024$, with an empirically estimated travel time (from $20$ R$_{\odot}$) of $44.23$ h. 

\subsection{Classification of the AR evolution}

\begin{figure}[h]
	\begin{center}
		\includegraphics[width=\textwidth]{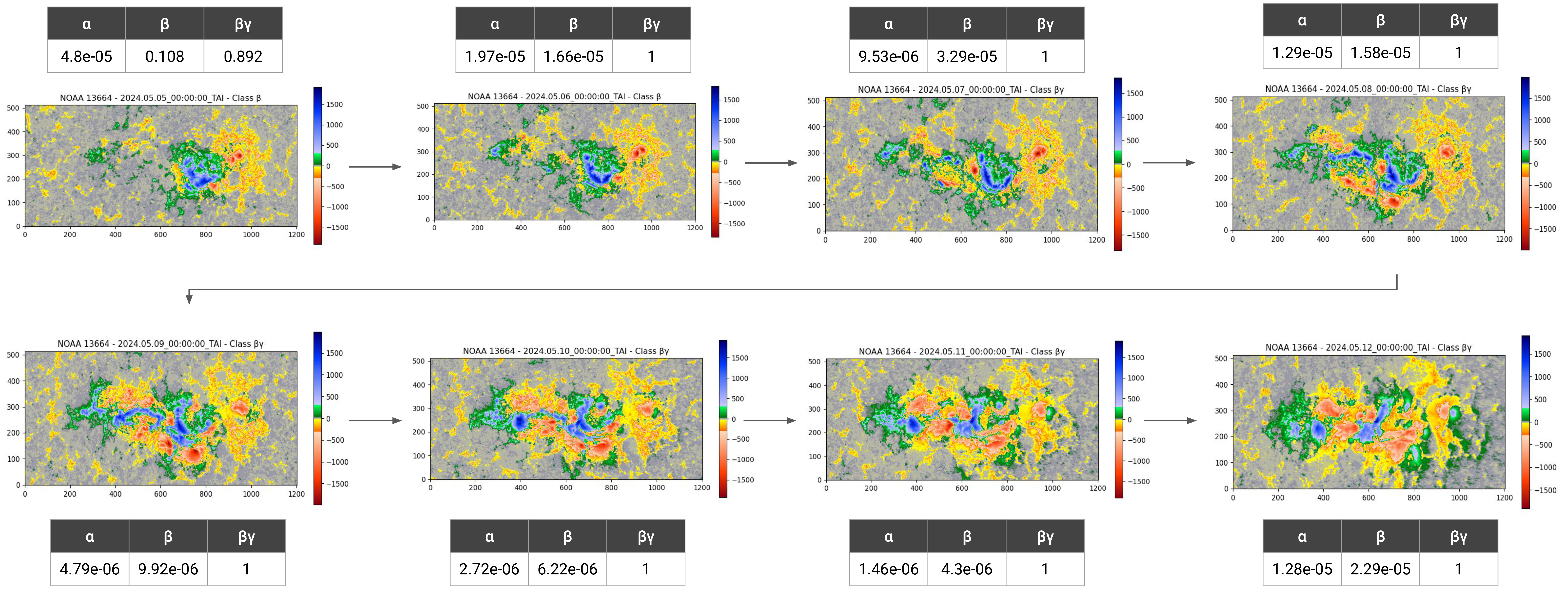}
	\end{center}
	\caption{Classification of AR13664 provided by a Data-Efficient Image Transformer. The evolution of the magnetogram cut-outs is associated with the corresponding class probabilities.}
	\label{fig:classification}
\end{figure}

It is well-established that the likelihood of solar flare occurrence correlates with specific morphological properties of ARs and with their magnetic classification \citep{ahmed2013solar,georgoulis2021flare}. As a consequence, AR classification is the first issue to address in the flow of space weather prediction. AR13664 evolved from a $\beta-$class sunspot during the initial days of observations, to a $\beta$-$\gamma$ class on the following days. Its morphology then stabilized whenever it was within the telescope's field of view. We aimed at verifying whether AI was able to confirm this visual assessment by means of an automated deep-learning-based classification. Specifically, following the conclusions outlined in \citet{legnaro2024deep}, we considered a Vision Transformer (ViT) \citep{dosovitskiy2020image} trained using a transfer learning approach on the SOLAR-STORM1 dataset provided by the Space Environment Warning and AI Technology Interdisciplinary Innovation Working Group \citep{fang2019deep}. The optimized ViT was applied to $8$ HMI magnetograms associated to AR13664 from May 5, 2024 to May 12, 2024, and the corresponding classification flow over time is represented in Figure \ref{fig:classification}. This figure shows that deep learning is able to perceive a certain degree of morphological ambiguity in the May 5, 2024 magnetogram, whereby some trace of the $\beta-$ class component persists in the image. After that, our model had no longer doubt in classifying the sunspot as of $\beta$-$\gamma$ type.

\begin{figure}[h]
\begin{center}
    \includegraphics[width=\linewidth]{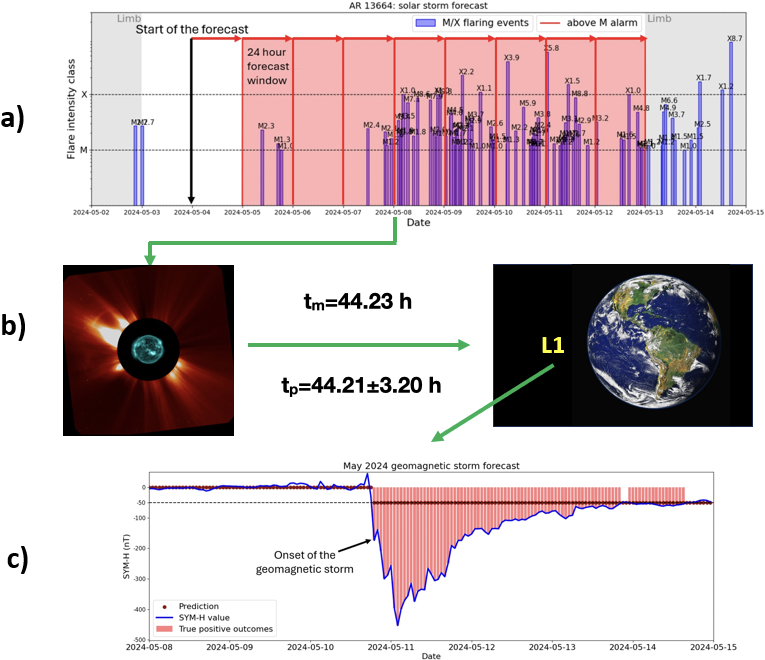} 
\end{center}
\caption{AI-driven forecasting results. The top panel enrols over time the alarms sounded by video-based deep learning. The numbers in the middle row compare the CME travel time predicted by the physics-driven neural network ($t_p$) with the measured one ($t_m$). The panel at the bottom enrols over time the prediction of the storm's magnetic impact assessed via forecast of SYM-H values.}\label{fig:prediction}
\end{figure}

After this preliminary classification step, the subsequent analysis combined pioneering AI-based methodologies with remote sensing and in-situ data to predict the entire sequence of space weather events, from flare occurrences to CME arrival time and the geomagnetic impact of the solar storm. The results of this pipeline of prediction approaches are contained in Figure \ref{fig:prediction}.

\subsection{Flare forecasting}

We addressed a fully data-driven flare forecasting task by means of a video-based deep learning network made by a set of convolutional neural networks (CNNs) working in parallel (one CNN for each video frame) and a long-short term memory (LSTM) network realizing prediction by taking as input the feature sequences extracted by the CNNs \citep{donahue2015long}. We constructed 24-hour-long videos of AR13664 cutouts (with a cadence of $36$ minutes) and, at each time step, the algorithm was trained by means of a score-oriented optimization process \citep{marchetti2022score} in order to predict the occurrence of solar flares above M class in the next $24$ hours. The training and validation sets were generated using the HMI archive of magnetograms in the time range between 2012 September 14 and 2017 September 30, and an accurate balancing of the sets was realized by applying the algorithm introduced in \citet{guastavino2022implementation}. The top panel of Figure \ref{fig:prediction} shows the alarms sounded by our AI as enrolled over time: the first prediction was made on 2024-05-04 for the next 24 hours and the last prediction was made on 2024-05-12 for the next 24 hours (after that date AR13664 was out the field of view of HMI). The algorithm sent warnings of the M+ flares from 2024-05-05 to 2024-05-13 by correctly predicting the most intense M+ and X+ flares. However, the alarm sounded in the window between 2027-05-06 and 2024-05-07 corresponded to a false positive, since no flare occurred that day.

\subsection{Prediction of the CME travel time}

As said, the solar flares predicted as described in the above sub-section originated $5$ CMEs in succession, the last one being able to ingest all the previous ones into a single impressive cloud. In this study, we performed the characterization of this last CME by means of a neural network whose loss function was designed in order to account for a deterministic modeling of its dynamic relying on the analytical solution of the drag-based equation \citep{2023ApJ...954..151G}. The CME dynamical and morphological parameters used as input in this physics-driven AI method were obtained by applying the cone model \citep{2002JGRA..107.1223Z,2004JGRA..109.3109X} to a set of LASCO coronagraphic images. This allowed the inference of the velocity of the CME at $20$ R$_{\odot}$, its trajectory, and its angular width. An estimate of the CME mass was also provided using the method described in \citet{2000ApJ...534..456V,2010ApJ...722.1522V}. Finally, the density and speed of the solar wind in which the CME propagates were retrieved from a semi-empirical model of the ecliptic distribution of the solar wind, which was obtained by combining in-situ plasma measurements with the analytical model of the interplanetary magnetic field (IMF) structure by \citet{1958ApJ...128..664P}. These estimated parameters are included in Table \ref{tab:CME}. The last row of the table contains the travel time predicted by means of our physics-driven AI model, which is impressively close to the observed one (note that the uncertainty in the predicted time was computed by means of an ensemble learning approach for optimization). The predicted travel time is reported as well in Figure \ref{fig:prediction}, middle panel, where it is compared with the observed one.

\begin{table}[]
\centering
\caption{List of CME and solar wind input parameters, the actual travel time and the predicted travel time by the physics-driven AI method.}
\begin{tabular}{llll}
 Parameter & Notation & Unity & Value \\ \hline \hline
\multicolumn{4}{c}{Input parameters} \\ \hline \hline
CME initial speed at $r_0 = 20$ $R_{\odot}$ & $v_0$ & km/s & $v_0 = 2598.65$    \\
CME mass & $ m$ &  g & $m=1.37\times10^{16}$   \\
CME angular width  & $\phi$ & degree & $\phi=166.54$\\
Solar wind number density & $\rho$ & cm$^{-3}$ & $\rho=3$ \\
Solar wind speed   & $w$ & km/s & $w=500$ \\
\hline \hline
\multicolumn{4}{c}{Actual label} \\ \hline \hline
CME time of eruption at $r_0$ & $t_0$ & Date & $t_0 = $ 2024-05-08 T22:01 \\
CME Time of Arrival & $t_f$ & Date & $t_f$ = 2024-05-10 T18:15  \\
CME Travel time & $t$ & h &  $t = 44.23$ \\ \hline \hline
\multicolumn{4}{c}{Predicted label} \\ \hline \hline
Predicted CME Travel time & $\hat{t}$ & s &  $\hat{t} = 44.21 \pm 3.20$ \\ 
\end{tabular} 
\label{tab:CME}
\end{table}

\subsection{Forecast of the geomagnetic storm}

As for the forecast of the geomagnetic storm the predictions provided by the machine learning approach based on in-situ data developed by \citet{2024ApJ...971...94G} were also significantly accurate. In this case, the input parameters were represented by the 24-hour time series of features derived from situ measurements and the SYM-H index, which measures the symmetric portion of the horizontal component magnetic field near the equator \citep{wanliss2006high}. Such input features revealed predictive capabilities through different analysis of feature selection methods conducted in \citet{2024ApJ...971...94G} and in \citet{camattari2024classifier}. In detail, this set of features contains:
\begin{itemize}
\item The magnitude of the magnetic field vector $\mathbf{B}$ and its z-component $B_z$.
\item The magnitude of the solar wind velocity vector $\mathbf{V}$ and its x-component $V_x$.
\item The solar wind temperature $T$.
\item The magnetic helicity $H_{m}$.
\item The kinetic $E_{k}$ and magnetic energy $E_{m}$, and the total energy $E=E_{k}+E_{m}$.
\end{itemize}
For a comprehensive discussion on how these quantities were estimated in the solar wind, the reader is referred to \citet{2019ApJ...885..120T} and \citet{2020ApJ...896..149T}. As illustrated in Figure \ref{fig:prediction}, bottom panel, AI is able to accurately predict not only the onset but also the whole recovery phase $1$ hour in advance. As expected, some uncertainty occurred only in the last part of the recovery phase, when the SYM-H fluctuates around $-50$ nT. To obtain this prediction we have applied once again the LSTM network, with a score-oriented loss function for the automated optimization of the True Skill Statistic (TSS) score.

\section{Discussion}
This work highlights in one shot the notable potential of AI in revolutionizing space weather forecasting when the prediction task involves the whole chain of events triggered by solar flares and including CME characterization and the prediction of the geomagnetic impact. The integration of advanced data-driven AI techniques with physics models, has proven instrumental in improving the accuracy of prediction.

The case study of the May $2024$ superstorm illustrates the profound capabilities of AI. In fact, the efficiency with which a ViT is able to recognize the morphological properties of AR13664 should mainly be attributed to the high degree of sophistication with which the training process took place. Indeed, the combination of data augmentation and transfer learning seems to have made it possible to overcome the classification limitations imposed by the imbalance of the historical archive, to the point where ViT even seems able to feel the occurrence of transition morphologies between one configuration and the next one. These impressive performances of deep learning are confirmed by the fact that a hybrid architecture made of CNNs and an LSTM network was able to forecast the early May 2024 flaring storm with a remarkable degree of reliability. This prediction effectiveness enforces what was already shown by \citet{benvenuto2020machine} and theoretically explained by \citet{guastavino2022bad} by means of standard machine learning approaches, i.e., that forecasting flaring storms is easier than identifying in advance possible isolated events.

The extreme solar events of May 2024, culminating in a geomagnetic storm of unprecedented intensity, offered a unique opportunity to leverage AI not only as a predictive tool but also as a means to extract fundamental insights into the physics of CME dynamics. The accurate prediction of the CME travel time during this event, with an error margin of less than one minute, underscores the potential of AI-driven approaches to decode the complex interplay of forces acting on CMEs as they propagate through the heliosphere.

Further, the exceptional velocity of the May 8 CME, exceeding 1000 km/s, coupled with its interaction with multiple preceding ejections, highlights the importance of understanding CME-CME interactions in shaping the transit dynamics. The ability of AI to predict the combined CME travel time provides a reverse-engineering framework to examine the underlying processes, such as cannibalism and magnetic reconnection, that occurred as these ejections merged into a single, highly energetic structure. These interactions not only influence the final velocity and morphology of the CME but also encode information about the heliospheric environment, including density gradients and magnetic field configurations along the CME trajectory, as well as the drag forces and accelerative effects exerted by the solar wind. 

In the context of the May 2024 event, the remarkable accuracy of the predicted transit time reflects the robustness of the physics-driven AI model employed, which incorporates a deterministic drag-based framework. The methodology captures the complex interplay of aerodynamic drag, solar wind interactions, and interplanetary dynamics, providing a comprehensive framework to describe the CME’s propagation. These insights may enhance our understanding of heliospheric dynamics during periods of high solar activity.

Upon reaching Earth’s magnetosphere, the CME’s interaction triggered a geomagnetic storm classified as G5, marking it as the second most severe disturbance recorded in the space era. The tight correlation between the AI-predicted CME arrival time and the onset of the storm provides an indirect probe into the temporal and spatial scales of energy transfer from the CME to the magnetosphere. In particular, the rapid onset of geomagnetic activity following the CME’s arrival is consistent with an efficient coupling mechanism, likely mediated by the favorable orientation of the interplanetary magnetic field (IMF). By analyzing the AI-predicted parameters in conjunction with in-situ measurements of the IMF and solar wind plasma, it would be in principle possible to reconstruct the CME’s magnetic topology and assess its role in driving magnetospheric dynamics.

As a conclusion, the May 2024 event also underscores the broader implications of AI-driven reverse engineering for space weather science. The ability to predict CME travel times with such precision suggests that AI can furthermore serve as a diagnostic tool for testing and refining existing models of CME propagation. The integration of AI with physics-based modeling provides indeed a pathway for disentangling the relative contributions of different forces acting on CMEs, such as solar wind drag, magnetic pressure, and interactions with interplanetary structures. Furthermore, the success of AI in predicting geomagnetic storm onset and recovery phases one hour in advance highlights its potential to illuminate the temporal evolution of solar-terrestrial coupling processes. This event exemplifies the dual role of AI in space weather research: not only does it improve operational forecasting, but it also offers a new lens through which to explore the fundamental physics of solar and heliospheric dynamics. These advancements could significantly reduce the socioeconomic impacts of severe space weather events by enabling earlier and more precise warnings. 

Future studies could build upon these findings by applying similar methodologies to other extreme events, thereby expanding our understanding of the heliosphere’s behavior during periods of intense solar activity. The May 2024 superstorm stands as a testament to the transformative potential of AI in bridging the gap between predictive capabilities and physical insight.

\section{Methods}
The analysis performed in this study is built upon novel AI-based technologies that have been designed and optimized in order to account for the specific aspects of the prediction/classification tasks associated to each phase of the chain of events characterizing the May 2024 superstorm. 

\subsection{Vision Transformer for active region classification}
\begin{figure}
\begin{center}
\includegraphics[width=0.75\textwidth]{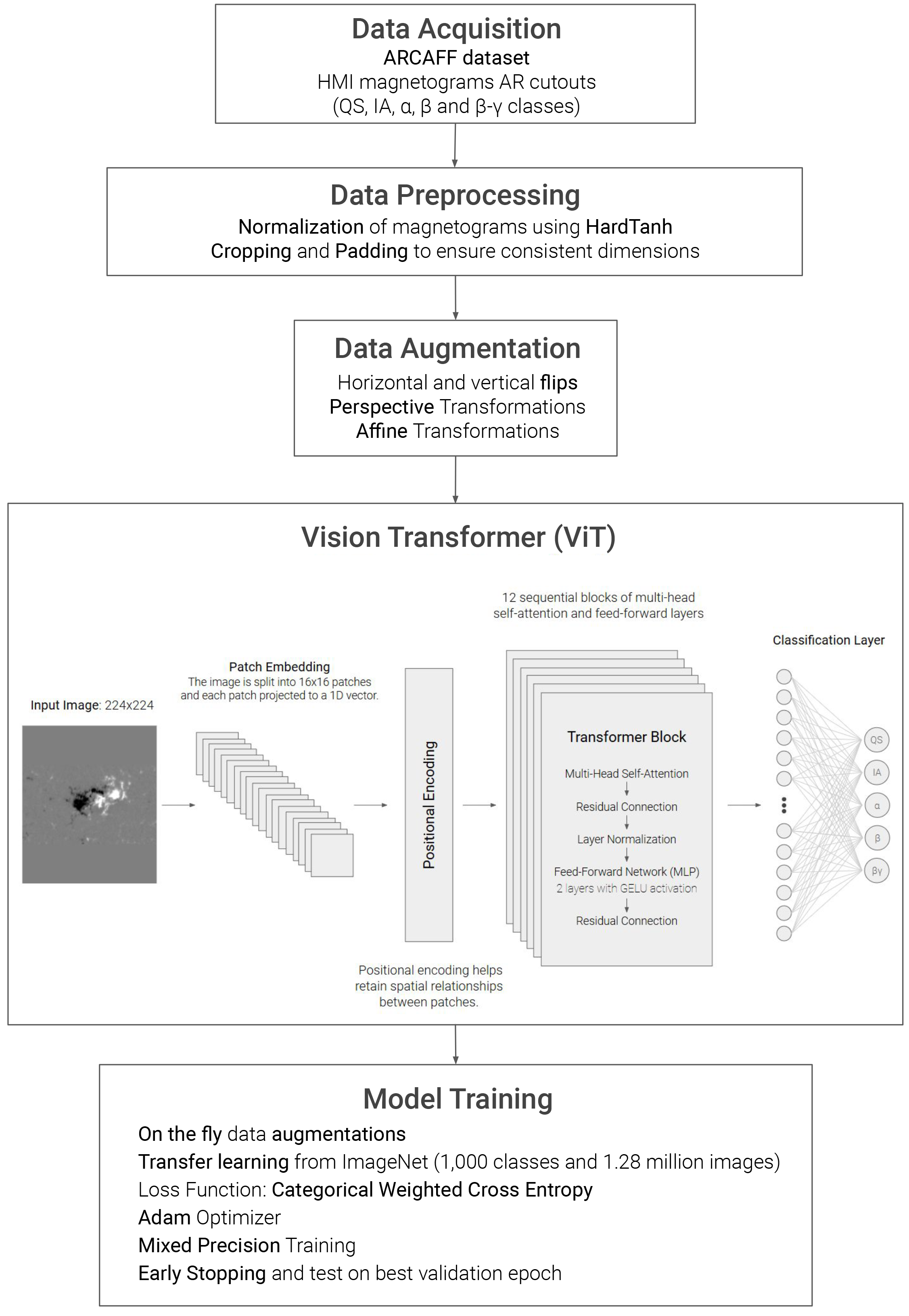}
\caption{The figure illustrates the pipeline for classifying active region (AR) magnetograms using a Vision Transformer (ViT) neural network. The process begins with the considered dataset, consisting of AR magnetogram cutouts. Low-occurring classes are either dropped or merged. Preprocessing involves normalization with HardTanh function (so the values are in $[-1, 1]$) and ensuring consistent dimensions through cropping and padding. Data augmentation is performed by applying flips, perspective, and affine transformations. The ViT model processes the input as patcsh embeddings with positional encoding through 12 transformer blocks to a classification layer. Training includes transfer learning from ImageNet, on-the-fly augmentations, categorical weighted cross-entropy loss, mixed precision training, and early stopping based on validation performance.}
\label{fig:pipeline}
\end{center}
\end{figure}

To characterize the morphological evolution of AR13664, we utilized the computational pipeline illustrated in Figure \ref{fig:pipeline}. This consists of the following.
\begin{enumerate}
    \item A pre-processing step made of normalization, cropping, and padding for ensuring input dimensions consistency.
    \item A data augmentation step for expanding the training set.
    \item A temporal data splitting is applied to the available data in order to generate training, validation and test sets, and to allow cross-validation.
    \item A ViT model architecture is selected.
    \item A training step that exploits on-the-fly data augmentation and transfer learning from the ImageNet dataset.
\end{enumerate}
More details on the training process can be found in \citet{legnaro2024deep}.
Here we applied the ViT to the magnetograms associated with AR13664. Specifically, the ViT diagram in Figure \ref{fig:pipeline} illustrates the flow of data through the main components of the ViT model, starting from the input image to the final output logits for classification. 
The input is a grayscale 224x224 image, which is then passed to the Patch Embedding layer, where the input image is split into non-overlapping patches of size 16x16. 
Each patch is flattened into a 1D vector and then projected into a 384-dimensional embedding through a linear layer. 
The resulting sequence consists of 196 patch embeddings, which becomes 197 with the addition of a classification token, a special learnable embedding that serves as a summary representation of the input data.
To retain the spatial information of patches, a positional encoding vector is added to each patch embedding element-wise. 
Then, the data go through 12 Transformer blocks, each one containing a Multi-head Self Attention (MSA) module, a Residual Connection Skip (RCS) module, a Layer Normalization (LN) module, and a Feed-Forward Network.  
These blocks are repeated 12 times, producing an output sequence of $197 \times 384$.
The final output sequence retains the classification token with its updated representation after passing through the Transformer blocks.
Finally, the token (shape $1 \times 384$) is extracted and passed through a fully connected (linear) layer to project it to the output classes ($1 \times 5$ in this case), and the model outputs a vector of logits (shape $1 \times 5$), representing the predicted class probabilities for the 5 target classes.

\subsection{Deep learning for flare forecasting}
\begin{figure}
\begin{center}
\includegraphics[width=14.cm]{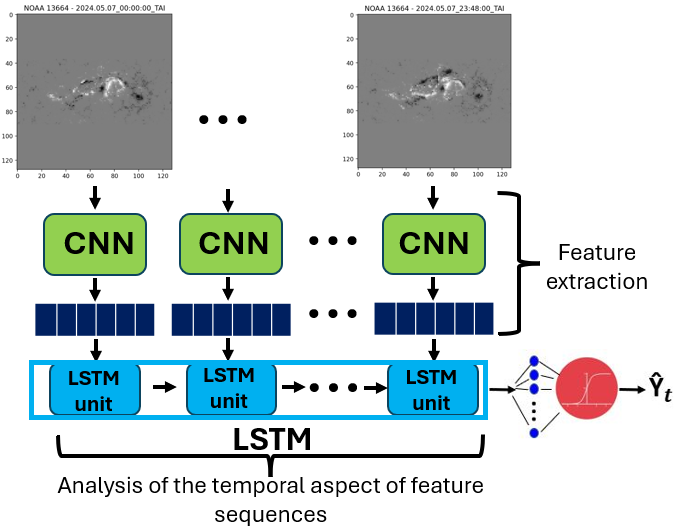}
\caption{Video-based deep neural network architecture for flare forecasting.}
\label{fig:video-based}
\end{center}
\end{figure}

\citet{guastavino2022implementation} introduced a novel methodology for flare forecasting, which is a video-based deep learning model specifically designed by merging convolutional and long short-term memory architectures (see Figure \ref{fig:video-based}). This Long-term Recurrent Convolutional Network (LRCN) \citep{yu2019review} is made of a sequence of 
\begin{itemize}
\item A 7x7 convolutional layer.
\item A 2x2 max-pooling layer.
\item A 5×5 convolutional layer.
\item A 2×2 max-pooling
layer.
\item A 3×3 convolutional layer.
\item A 2×2 max-pooling layer.
\item A 3×3 convolutional layer.
\item A 2×2 max-pooling
layer.
\item A final dense layer of 64 units (where dropout was applied with a fraction of 0.1 input units dropped).
\end{itemize}
A Rectified Linear Unit (ReLU) is applied in all cases. All convolutional layers are made of 32 units with height and width
strides equal to 2 (these same parameters have been set to 1 for max-pooling). Further, we applied L2-regularization and a standardization process to the convolutional layer outcomes, and a flattening process before applying
the dense layer. In this design, the HMI video is considered as a time series of frames, each one given as input to each CNN of the pipeline in parallel. The output of this parallelized process, which is made of 40 vectors, each one made of 64 features, is fed into the 50-unit LSTM that performs classification (also for LSTM dropout was applied). Finally, a dense sigmoid unit is used to map the output of the LSTM to the range [0,1], enabling the model to perform binary classification.

\subsection{A physics-driven neural network for CME characterization}

\begin{figure}
\begin{center}
\includegraphics[width=14.cm]{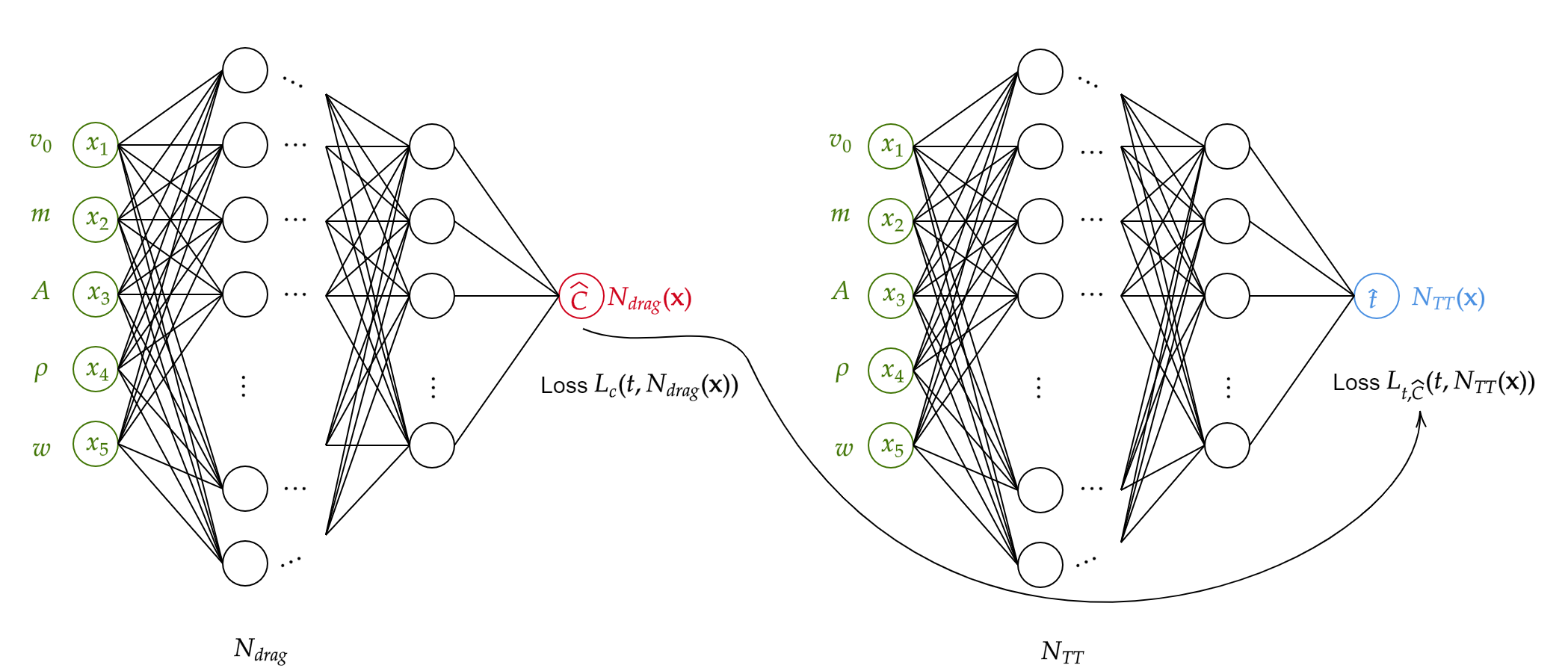}
\includegraphics[width=14.cm]{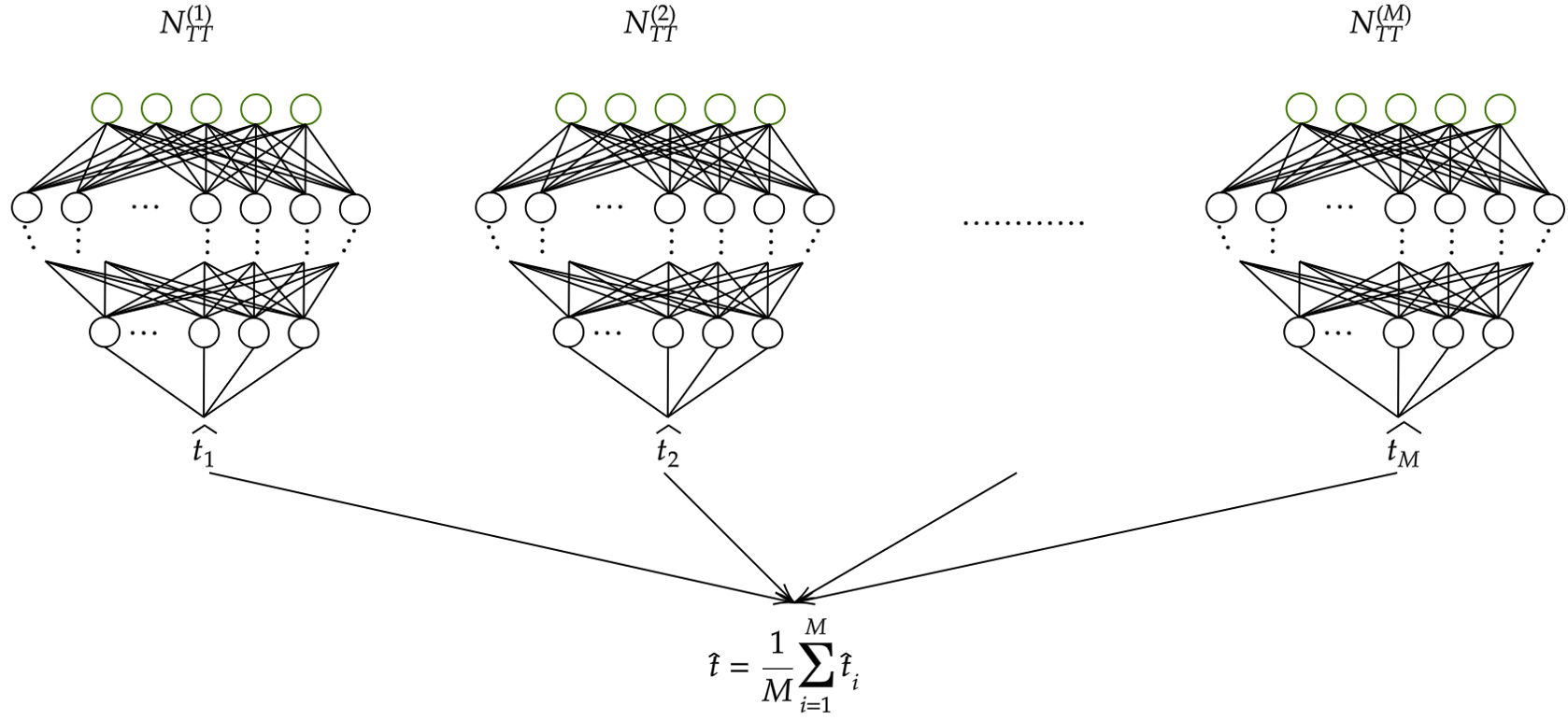}
\caption{The physics-driven AI method for the CME characterization. Top panel: the cascade of two neural networks that allows the computation of the drag parameter and of the CME travel time. Bottom panel: scheme of the ensemble physics-driven AI algorithm.}
\label{fig:NN-drag}
\end{center}
\end{figure}

\citet{2023ApJ...954..151G} presented an unprecedented approach to predicting the CME travel time using an hybrid physics-driven AI method. That approach (see Figure \ref{fig:NN-drag}, top panel) encoded a deterministic drag-based model (DBM) into the training process of a cascade of two neural networks. The DBM, a well-known kinematic model, simplifies the interaction between CMEs and the solar wind by modeling drag acceleration as a quadratic function of the relative speed between the CME and the solar wind. Although DBMs have limitations due to their simplified assumptions, their computational efficiency and limited input parameters make them suitable for this kind of studies.
In greater detail, the first neural network (denoted as $N_{drag}$) is trained on an historical dataset to estimate the unknown drag parameter that appears in the DBM equation by taking in input initial CME speed ($v_0$), mass ($m$), impact area ($A$), solar wind density ($\rho$) and speed ($w$); then, the second neural network (denoted as $N_{TT}$) used the same input parameters as the first one
to predict the CME travel time. The loss function for the first neural network $L_C$ is fully model-driven, minimizing the discrepancy between the predicted distance by the DBM model with the estimated drag-coefficient and the actual distance Sun-Earth (around 1 Astronomical Unit AU). It is defined as follows
\begin{equation}\label{eq:loss-1}
L_c(t,N_{drag}(x)) = (r(t,N_{drag}(x))-1)^2 \,
\end{equation}
where 
$r$ represents an approximation of the analytical solution of the drag-based model, $x$ is the vector input made by the following components $v_0,m,A,\rho,w$ and $t$ is the measured travel time.
The loss function of the second neural network $L_{t,\hat{C}}$ combines both data- and physics-driven components, i.e. a mean square error between the predicted and the actual travel times and the DBM-inspired loss function used in the first neural network. It is expressed as follows
\begin{equation}
     L_{t,\hat{C}}(t,N_{TT}(x)) =  \lambda (t - N_{TT}(x) )^2  +  (1 -\lambda) (r(N_{TT}(x),\hat{C})-1)^2, 
\end{equation}
where $\lambda$ balances the two components and $\hat{C}$ is the estimated drag parameter from the first neural network, i.e. $N_{drag}(x)$.  
The neural networks were trained multiple times using different splits of an archive of CME events spanning from 1993 to 2018. To predict the CME’s travel time, several of the trained networks were applied within an ensemble learning framework (see Figure \ref{fig:NN-drag}, bottom panel), with the overall predicted travel time computed as the average of the individual network predictions. This ensemble approach also enables the estimation of uncertainty in the predicted travel time.

\subsection{A neural network for predicting the geomagnetic impact of the storm}

 \citet{2024ApJ...971...94G} advanced the capabilities for forecasting geomagnetic storms by leveraging machine learning techniques. Specifically, a LSTM network, known for its effectiveness in processing sequential data, was integrated in the analysis of in-situ measurements of the solar wind parameters acquired at L$1$. 

The model architecture consisted of an LSTM layer with 72 units, followed by a dense layer with 64 units, and a final dense layer with a single unit using a sigmoid activation function. This configuration was designed to predict the probability that the SYM-H index drops below $-50$ nT within the next hour. The network was trained and validated on a historical data set, in which the in-situ measurements were acquired by Wind, in the time range between January 1, 2005, and December 31, 2023. 
Specifically, the training and validation split preserved the significant imbalance of the dataset (with approximately 1.95\% positive samples), and a suitable Score-Oriented Loss (SOL) function was used to optimize the True Skill Statistic (TSS) during the network's training process \citep{marchetti2022score}. In order to prevent overfitting, the network was trained over $1000$ epochs and an early stopping condition was imposed when no improvement in the validation loss was obtained after $100$ epochs.

In addition, for the first time, the energy budget of the solar wind and the magnetic helicity were also considered as input into the neural network prediction model, in order to enhance prediction accuracy. Those innovations enabled more accurate forecasts of geomagnetic storm severity, with a particular emphasis on properly predicting both the onset (in $98\%$ of cases) and the recovery phases of geomagnetic disturbances, as shown in \cite{2024ApJ...971...94G}.

\section*{Acknowledgements}
SG acknowledges the financial support of the Programma Operativo Nazionale (PON) ``Ricerca e Innovazione'' 2014-2020. EL, AMM, and MP acknowledge the support of the HORIZON Europe ARCAFF Project, Grant No. 101082164; AMM, DT and MP acknowledge the support of the Fondazione Compagnia di San Paolo within the framework of the Artificial Intelligence Call for Proposals AIxtreme project (ID Rol: 71708). SG, MP and AMM are also grateful to the Gruppo Nazionale per il Calcolo Scientifico
- Istituto Nazionale di Alta Matematica (GNCS -
INdAM).

\bibliographystyle{aa}
\bibliography{reference-list}

\end{document}